# Beam-controlled spectral-selective architecture with planar polydimethylsiloxane/metal-films for all-day radiative cooling

Lyu Zhou[1,*], Haomin Song[1,2,*], Jianwei Liang[2,*], Matthew Singer[1], Ming Zhou[3], Edgars Stegenburgs[2], Nan Zhang[1], Chen Xu[4], Tien Khee Ng[2], Zongfu Yu[3,†], Boon Ooi[2,†], Qiaoqiang Gan[1,†]

1 *Department of Electrical Engineering, The State University of New York at Buffalo, Buffalo, NY 14260, USA.*

2 *KAUST Nanophotonics Lab, King Abdullah University of Science and Technology, Thuwal 23955-6900, Saudi Arabia*

3 *Department of Electrical and Computer Engineering, University of Wisconsin, Madison, Wisconsin 53705, USA*

4 *School of Life Information Science and Instrument Engineering, Hangzhou Dianzi University, Hangzhou 310018, Zhejiang Province, China*

\* These authors contribute to this work.

† Email: zyu54@wisc.edu; boon.ooi@kaust.edu.sa; qqgan@buffalo.edu

**Abstract**- Radiative cooling is a passive cooling strategy with zero consumption of electricity. Although this technology can work well during optimal atmospheric conditions at nighttime, it is essential to achieve efficient radiative cooling during daytime when peak cooling demand actually occurs. In this article, we report an inexpensive planar polydimethylsiloxane (PDMS)/metal thermal emitter, i.e., a thin film structure fabricated using fast solution coating process that is scalable for large area manufacturing. By manipulating the beaming effect of the thermal radiation, temperature reduction of 9.5 °C and 11.0 °C were demonstrated in the laboratory and out-door environment, respectively. In addition, a spectral-selective solar shelter architecture was designed and implemented to suppress the solar input during the daytime. Due to the enhanced directionality of the thermal emission, the dependence of the radiative cooling performance on the surrounding environment was minimized. Out-door experiments were performed in Buffalo NY, realizing continuous all-day radiative cooling with an average power of ~120 W/m$^2$ on a typical clear sunny day at Northern United States latitudes. This practical passive cooling strategy that cools without any electricity input could have a significant impact on global energy consumption.

**Introduction**

Air conditioning is a significant end-use of energy and a major driver of global peak electricity demand. For instance, air conditioning consumes ~15% of the primary energy used by buildings in the United States [1,2] and a shocking 70% of total electricity consumption in some tropical countries (e.g. Saudi Arabia [3]). Therefore, a passive cooling strategy that cools without any electricity input could have a significant impact on global energy consumption. The Earth's atmosphere has a transparent window for electromagnetic (EM) waves between 8-13 μm, corresponding to the peak thermal radiation spectral range of terrestrial objects at typical ambient temperature (e.g. ~20 °C to 45 °C). This transparent window is a cooling channel, through which a thermal body on the Earth's surface can radiate heat into the cold outer space. This mechanism has been widely used for nocturnal ice making in India before the invention of refrigerator [4]. In the studies of heat management in modern buildings, color and material properties of roofs and windows have been exploited for radiative cooling for decades (e.g. ref. [5]), which can help to reduce the heat gain through the roof during daytime and to cool the room/house more rapidly after sunset. However, most conventional radiative cooling technologies only work at nighttime since the solar heating is dominant during the daytime. To realize the envisioned all-day continuous cooling, it is essential to achieve efficient radiative cooling during daytime when peak cooling demand actually occurs [6-8].

Recently, record-breaking daytime radiative cooling strategies were demonstrated experimentally using various thermal photonic structures. For instance, a planar multi-layered photonic device for daytime radiative cooling has been reported [9], which reduced the temperature by 4.9 °C below ambient temperature. However, the multi-layered thermal photonic structure requires nanometer level accuracy in thickness controllability, imposing a cost barrier in practical daytime radiative cooling. To overcome this cost barrier, hybrid metamaterial structures fabricated using roll-to-roll processes were reported with the cooling power of ~100W/$m^2$ during a sunny day with clear sky [10]. In addition, photonic structures with high visible-to-IR transparency and strong thermal emissivity were also reported [11-13], which is particularly useful for improved operation of solar panels: i.e., the transparency in visible-to-IR regime will allow efficient solar light harvesting and the strong thermal emission will cool down the temperature of solar panels which is helpful to increase the efficiency and extend the device lifetime [14,15]. These pioneering works demonstrated the potential to realize daytime radiative cooling with no electricity consumption. This technology can be used to assist the climate control in buildings, saving a significant portion of energy usage [16-19]. Therefore, enhanced radiative cooling technology represents a new research topic with significant impact on the energy sustainability. However, the relatively expensive fabrication imposed a cost barrier for practical application of these daytime cooling strategies. In recent years, it becomes an emerging topic to improve the thermal management for the system and enhance the performance-to-cost ratio for daytime radiative cooling [9-14, 20-27].

In this article, we report an inexpensive planar PDMS/metal thermal emitter thin film structure that is useful for efficient radiative cooling applications over large areas. By manipulating the beaming effect of the thermal radiation, a temperature reduction of 9.5 °C was demonstrated using liquid nitrogen as the cold source in the laboratory environment. In addition, since the usual thermal emission of the planar thermal emitter is omnidirectional, the radiative cooling performance is heavily dependent on the surrounding environment (i.e., the access to the open clear sky). Due to the enhance directionality of the thermal emission, the dependence of the radiative cooling performance on the surrounding environment was minimized. In addition, a spectral-selective solar shelter architecture was designed and implemented to suppress the solar input during the daytime. Out-door experiments were performed in Buffalo NY, realizing continuous all-day radiative cooling with the best temperature reduction of 11.0 °C and an average cooling power of ~120 W/$m^2$ on a typical clear sunny day at Northern United States latitudes.

**Design and Development of planar PDMS/metal thin film thermal emitters**

PDMS is a promising inexpensive material for daytime radiative cooling due to its transparency in visible regime and strong thermal emissivity in the mid-infrared regime. In a recent report [28], 2.5-μm ~ 16.7-μm thick PDMS films were spin-coated on Au films. Based on their optical absorption properties, the application for radiative cooling was proposed using numerical modeling. Here we will reveal that this accurate thickness control is unnecessary for high performance radiative cooling in practice. Instead, we propose a simple planar PDMS/metal (Al or Ag) film structure to realize an inexpensive thermal emitter for radiative cooling, as illustrated in **Fig. 1A**. For a 150-μm-thick PDMS film, the optical absorption in visible to near IR spectrum domain is relatively weak, which is desired for daytime radiative cooling (see the inset of **Fig. 1B** for the measured data). According to Kirchhoff's law, the absorption of the emitter corresponds to its emissivity. Importantly, one can see from measured data shown in Fig. 1B (spheres) that its optical absorption/emissivity in the mid-infrared spectral range is strong, agreeing well with the numerical modeling (see the solid curve, the optical data of PDMS are shown in *Fig. S1* in the supplementary material). To reveal the thickness-dependence of this type of planar PDMS/metal thermal emitter, the absorption spectra of the thin film system was modeled as the function of the PDMS thickness, as shown in **Fig. 1C**. One can see obvious interference phenomenon in the wavelength range of 15-25 μm. The absorption/emissivity in the wavelength range of 8-13 μm is close to unity when the PDMS thickness is beyond 100 μm. Therefore, this structure is tolerant of large roughness in a PDMS film with thickness over 100 μm, which is convenient for inexpensive manufacturing over huge scales. We then employed a fast coating facility (RK K303 Multicoater, **Fig. 1D**) to fabricate the PDMS layer with controlled thickness (see processing details in *Note S1* in the supporting Information and *Movie S1*). The coating area in this inexpensive system is ~66 cm × 46 cm, demonstrating the potential of low cost and rapid manufacturing compared with multi-layered optical film deposition (e.g. ref. [14]). After thermal polymerization, a high quality PDMS layer on aluminum (Al) plate was obtained as shown in **Fig. 1E**. The thickness was characterized using a probe profilometer (Veeco, Dektak 8 advance development profiler), demonstrating the well-controlled film thickness of ~150 μm (**Fig. 1F**). Using this rapid process, we fabricated five samples and realized relatively stable control in thickness, as shown by the inset in Fig. 1F. Their optical absorption spectra (i.e., thermal emissivity) are characterized using an FTIR spectroscopy (Bruker Hyperon 1000). As shown in Fig. 1B, the measured data (spheres) agrees well with theoretical modeling (the solid curve). Importantly, the measured strong emissivity in the 8-13 μm spectral window (~94.6% of ideal blackbody radiation, see the shaded region in Fig. 1B) and weak absorption of solar energy (i.e., less than 10%) will enable the high performance daytime radiative cooling.

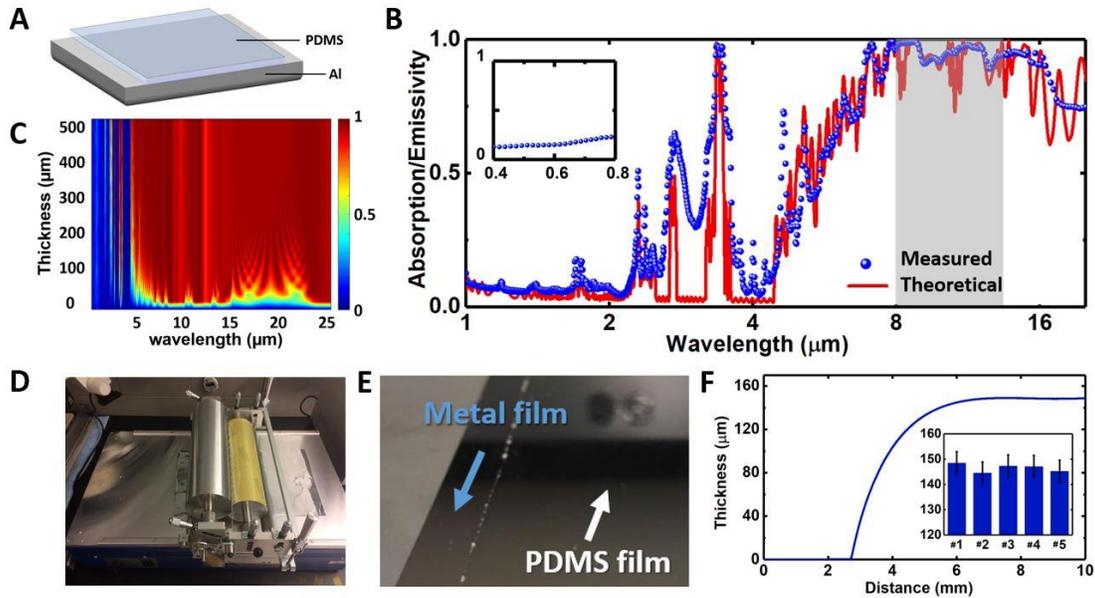

**Figure 1| PDMS/metal thin film thermal emitter.** (A) Schematic diagram of planar PDMS/metal thermal emitter. (B) Absorption/emissivity spectra of a planar PDMS/Al film with the thickness of 150 μm. Solid curve: numerical

modeling; Spheres: measured data. (C) Modeled absorption spectra of the planar PDMS/Al film as the function of the PDMS film thickness. (D) A photograph of the PDMS coating facility under operation. (E) A photograph of the edge of a coated PDMS film on an Al plate. (F) A cross-sectional profile of a PDMS film. Inset: Measured thicknesses of five samples.

**In-door experiment**

The schematic diagram of the in-door experimental setup is shown in **Fig. 2A**: We filled a bottom thermal isolating foam tank with liquid nitrogen (at 77K). At the bottom of this tank, we placed a black Al foil to absorb all thermal radiation. Its optical absorption spectrum from visible to mid-IR spectral range is shown in *Fig. S2*. In this experiment, the black aluminum foil in liquid nitrogen will act as a cold source. The PDMS/metal emitter was sealed by a polyethylene (PE) film in a thermal isolating foam container fixed at the top, facing down to the liquid nitrogen tank. The distance between the emitter and the cold source is ~1.5 meter (**Fig. 2B**). Three temperature probes were placed at different positions, as indicated by D1-D3 in Fig. 2A. To reveal the divergence of the thermal emission from the PDMS/Al emitter, we characterized the angle dependent absorption at the wavelength of 10 μm using the angle module in the FTIR system (Bruker A513 variable angle reflection accessory). One can see in **Fig. 2C** that the measured thermal emission of this thin film PDMS/Al film is approximately omnidirectional (left panel), agreeing very well with the numerical modeling result (right panel). Therefore, it is a technical challenge to collect the thermal radiation efficiently to optimize the radiative cooling performance. Beam control of the thermal emission (e.g. ref. [29]) is therefore of interest to address this limitation.

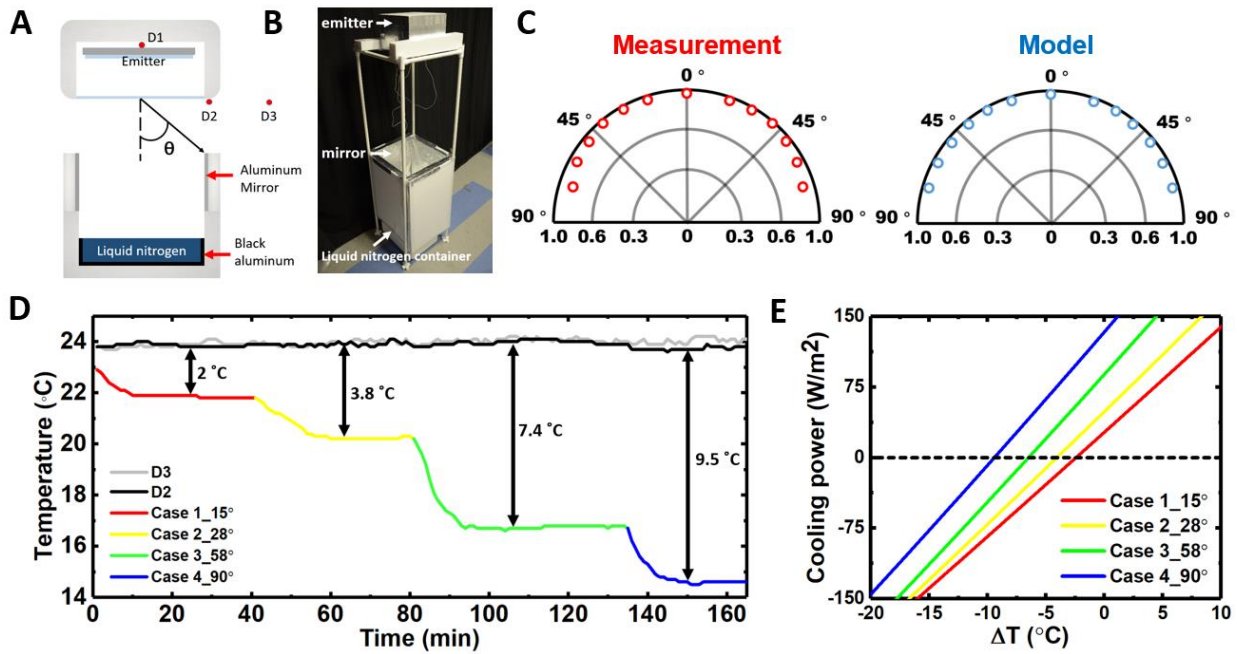

**Figure 2. In-door radiative cooling characterization using liquid nitrogen as the cold source.** (A) Schematic diagram of the experimental setup. (B) Photograph of the experimental setup. (C) Measured and modeled angle-dependent absorption distribution of the planar PDMS/Al cavity emitter at the wavelength of 10 μm. (D) Measured radiative cooling effect with different collection efficiencies (i.e., $\theta$ tuned from 15º to 90º). (E) Calculated cooling power of the 100-μm-thick PDMS/Al cavity emitter within different collection angles from 15º to 90º.

In this experiment, we partially or fully connected the output port of the emitter to the cold source using flat Al foils to form a rectangular waveguide tube for thermal emission. These Al mirrors were employed as the side wall to determine the collection angle, $\theta$, as illustrated in Fig. 2A. As a result, cooling effects at different collection angles were observed experimentally, as shown in **Fig. 2D**. When $\theta$ was tuned from 15º to 90º, a temperature difference from 2 ºC to 9.5 ºC was obtained, depending on the collection efficiency of the thermal radiation. It should be noted that the measured temperature at D2 and D3 are almost the same (black and gray curves, respectively), indicating that the temperature around the emitter box was not affected by the convection of liquid nitrogen. Therefore, the observed cooling effect was mainly introduced by wave-guided thermal radiation.

To interpret the observed in-door radiative cooling performance, we then analyze the cooling power at each collection angle. The net cooling power, $P_{net}$, is defined below:

$$P_{net} = P_{rad}(T_{dev}) - P_{amb}(T_{amb}) - P_{cold\ source}(T_{lN2}) - P_{nonrad}(T_{dev}, T_{amb}) \qquad (1).$$

Here, $P_{rad}$ is the output power of the PDMS/Al emitter:

$$P_{rad}(T_{dev}) = \int d\Omega \cos(\theta) \int d\lambda I_{BB}(T_{dev}) \varepsilon_{dev}(\lambda), \qquad (2)$$

where $T_{dev}$ is the temperature of the emitter; $\Omega$ is the solid angle; $I_{BB}(T) = \frac{2hc^2}{\lambda^5} \frac{1}{\exp\left(\frac{hc}{\lambda k_B T}\right)-1}$ is the spectral radiance of a blackbody at the temperature T; $\varepsilon_{dev}$ is the spectral emissivity/absorptivity of the PDMS/Al film; $h$ is the Planck's constant; $k_B$ is the Boltzmann constant; $c$ is the speed of light and $\lambda$ is the wavelength. The incident radiation powers from the ambient, $P_{amb}$, and from the cold source, $P_{cold\ source}$, are given by the following two equations, respectively:

$$P_{amb}(T_{amb}) = \int d\Omega \cos(\alpha) \int d\lambda I_{BB}(T_{amb}) \varepsilon_{air}(\lambda) \qquad (3)$$

$$P_{cold\ source} = \int d\Omega \cos(\beta) \int d\lambda I_{BB}(T_{lN2}) \varepsilon_0(\lambda) \qquad (4)$$

where $T_{amb}$ is the temperature of the ambient air; $T_{lN2}$ is the temperature of liquid nitrogen; $\varepsilon_{air}(\lambda)$ and $\varepsilon_0(\lambda)$ is the emissivity/absorptivity of air and the cold black aluminum foil, respectively. It should be noted that the calculated radiation power of the cold source ($P_{cold\ source}$) is negligible compared with other terms in Eq. (1) due to the very low temperature (i.e., 77K) (see detailed discussion in *Note S2* in the supporting material). The last term of Eq. (1), $P_{nonrad}(T_{dev}, T_{amb})$, is the nonradiative power loss because of convection and conduction, which is given by

$$P_{nonrad}(T_{dev}, T_{amb}) = q(T_{dev} - T_{amb}) \qquad (5).$$

Here $q$ is the conduction/convection efficiency. In the indoor experiment, the $q$ was defined as 10 W m$^{-2}$K$^{-1}$ (e.g. ref. [24]). Using these equations, the predicted cooling power of the system is plotted in **Fig. 2E** as the function of the temperature difference, $\Delta$T (i.e., the difference between the PDMS/Al emitter and the ambient). The intersection point at the cooling power of 0 W/m$^2$ indicates the achievable stabilized temperature difference. As the collection angle increases, the intersection points will shift to the left side, indicating the improved cooling performance. Here we calculated the cooling powers of the same emitter with four different collection angles of 15°, 28°, 58° and 90°, respectively, corresponding to the four experimental tests shown in Fig. 2D. The stabilized temperature differences at four intersection points of dashed line are -2.3 ºC, -3.7ºC, -6.5 ºC and -9.2 ºC, respectively, agreeing well with the measured results shown in Fig. 2D. Next, we continue to explore the radiative cooling performance by controlling the thermal emission angle in an out-door environment.

**Out-door experiment**

Radiative cooling was proposed to reduce the air-conditioning energy consumption. A clear sky is required for efficient thermal emission. However, how the surrounding environment near the emitter will affect the radiative cooling performance is a practical issue to justify the practical implementation of this passive cooling technology in urban areas. Before we perform the out-door experiment, it is necessary to analyze the angle dependence of the thermal emission first.

The emissivity/absorptivity of the atmosphere at the zenith angle, $\gamma$, can be described by [30]

$$\varepsilon_{air}(\gamma, \lambda) = 1 - [1 - \varepsilon_{air}(0, \lambda)]^{1/\cos\gamma} \tag{6}$$

where $[1 - \varepsilon_{air}(0, \lambda)]$ is equal to the modeled atmospheric transmission spectrum shown as the blue curve in **Fig. 3A** (data from MODTRAN®). Using this equation, we plot the angle dependent atmospheric transmission at the wavelength of 10 μm in **Fig. 3B**. One can see that the thermal emission to the real sky is no longer omnidirectional (in contrast to the indoor analysis shown in Fig. 2C). On the other hand, although the thermal emissivity reduces to ~75% at large emission angles, the structure still emits a significant part of its thermal energy at these angles. Therefore, if the thermal emission at these large angles is blocked in an outdoor environment by various surrounding architectures, the radiative cooling performance will be affected, as will be validated via experiment below.

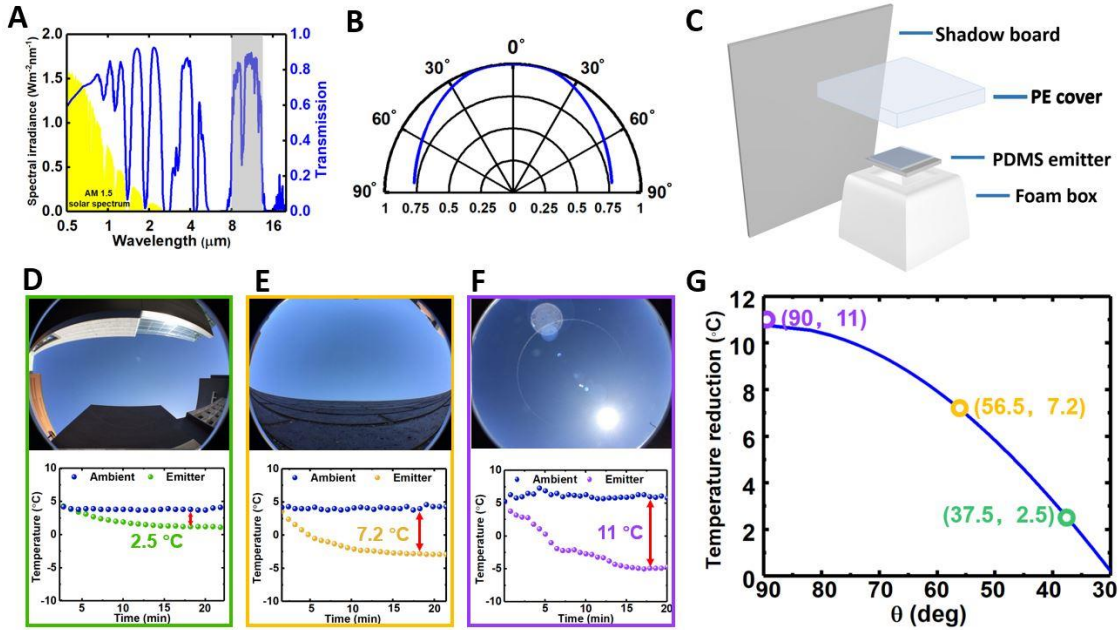

**Figure 3. Out-door radiative cooling test over different emission angle**. (A) Atmospheric transmission spectrum (blue curve) and the solar irradiation spectrum (yellow shaded region). (B) Modeled angle-dependent atmospheric transmission distribution at the wavelength of 10 μm. (C) Schematic of out-door radiative cooling test apparatus. (D)-(F) Measured temperature curves (lower panels) at different locations in UB (upper panels). (G) Calculated temperature reduction as the function of the collection angle (solid curve). Hollow dots: Measured data extracted from Figs. 3D~3F.

To reveal the environmental dependent cooling performance, we performed outdoor tests at three different places at UB campus from 11:00 am to 3:00 pm on February 28th, 2018 (with a clear sky and the relative humidity of ~60 %). The schematic diagram of our apparatus is illustrated in **Fig. 3C**. The planar PDMS/metal emitter was placed at the bottom of the high-density foam container sealed by the PE film. A foam board covered by highly reflective Al foils was placed next to the emitter container to serve as a shadow board (see *Fig. S3* in the supporting material). It can create a shadow to block the direct sun light illumination (see the spectrum plotted by the yellow region in Fig. 3A), especially within the period with peak solar input (inspired by ref. [13]). In particular, we employed a fish-eye lens (AMIR fisheye lens 180°) to demonstrate the access to the clear sky (see upper panels in **Figs. 3D-3F**, the lens was placed at the top of the shadow board). One can see that the access to the clear sky is limited when the emitter is surrounded by tall buildings (Figs. 3D and 3E). Large open spaces like parking lots are ideal for radiative cooling (Fig. 3F). As a result, we obtained the temperature reduction of 2.5 °C in Fig. 3D, 7.2°C in Fig. 3E and 11°C in Fig. 3F, respectively. Using Eqs. (1-6) (the temperature of the cold source is adapted to 3 K), we further modeled the temperature difference as the function of the collection angle (**Fig. 3G**). One can see that the estimated cooling performance (solid curve) agrees well with the experimental results extracted from Figs. 3D-3F (i.e., hollow dots). These results revealed a practical limitation to implement radiative cooling technology in urban areas: Although all buildings have the access to the clear sky on their roofs, the radiative cooling performance will be affected significantly by the surrounding architectures. To overcome this practical limitation, we will propose an improved system design using beaming effect of thermal radiation.

**A spectral-selective beaming architecture**

According to our calculation shown in Fig. 2E, the cooling power of the planar PDMS/Al system with a collection angle of 90° is ~120 W/m$^2$, corresponding to ~12% of the solar energy. However, as shown in the inset of Fig. 1B, the PDMS film still absorbs part of the solar irradiation in visible and near-infrared regime. In addition, the Al plates can also lead to the solar absorption at near-infrared wavelengths, which will considerably affect the cooling performance. Therefore, suppression of solar input is one of the most important technical issues for daytime cooling (e.g. ref. [9, 10, 13, 20, 22]). Here we introduce a spectral-selective absorber material (Bluetec coating solar collector, see the inset in **Fig. 4A**) in the design of the radiative cooling system. Its optical absorption spectrum is shown as the blue curve in **Fig. 4A**, with near-unity absorption of solar illumination and very high reflection in the mid-IR domain (the ultimate target is to reproduce the ideal absorption spectrum as shown by the green curve). As illustrated in the left panel in **Fig. 4B**, we employed this spectral selective film to design a tapered waveguide to serve as a beaming component. Therefore, most solar energy illuminated on its surface will be absorbed, while the thermal radiation from the planar thermal emitter can be reflected efficiently (right panel in **Fig. 4B**). Importantly, this material will not introduce much thermal emission back to the PDMS/Al emitter. At the center of the tapered waveguide, a smaller V-shaped shelter was introduced to block all normal incident solar light. As a result, the most important feature of this radiative cooling enhancement component is its beaming effect on the mid-IR radiation and the suppression of the solar input. As shown by the numerical modeling results in **Fig. 4C**, the incident solar light was absorbed by the shelter film at a wide range of angle (including the normal incident angle), resulting in a negligible solar input during daytime. By optimizing the taper angle, the mid-infrared wave can be collimated and be confined within a relatively small spatial angle (lower panel in **Fig. 4D**) compared to the planar system (upper panel in Fig. 4D) (see more modeling details in *Note* **S3**). Using this beam-controlled enhancement component (**Fig. 4E**), the impact of the surrounding environment in cooling performance can be greatly suppressed. To validate this prediction, we still placed the system at those three locations at UB (Fig. 3D-3F) and characterized their radiative cooling performance (two temperature probes were placed at different positions, as indicated by the red dots in the right panel in Fig. 2B). One can see from **Fig. 4F-4H** that the temperature reduction is very similar, confirming the independence on the surrounding architectures. More intriguingly, this beam-controlled spectral selective architecture can enable all-day radiative cooling, as will be discussed next.

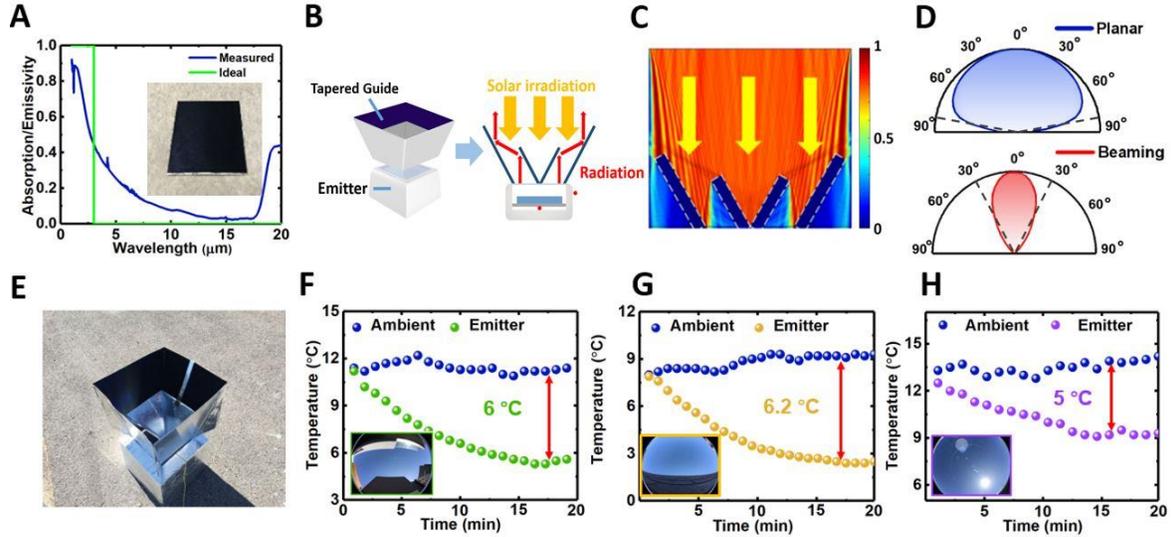

**Figure 4. Beaming effect and solar shelter for daytime cooling.** (A) Absorption spectra of an ideal selective absorber (the green line) and a commercial spectral selective absorber (the blue line), respectively. (B) Schematic diagram of the cooling system with tapered waveguide structure for thermal emission beam control and suppression of solar input. (C) Modeled beam propagation distribution with a normal incident solar light (at the wavelength of 500 nm) and (D) the output thermal beam propagation distributions for a planar system (upper panel) and a beaming system (lower panel) in mid-infrared wavelength region (at 10 μm). (E) The photograph of the beaming system. (F)-(H) Out-door experimental results at the same three locations at UB in Figs. 3D-3F.

**All-day continuous radiative cooling**

Finally, we performed continuous experiment at a co-author's backyard at Buffalo from 6:00 pm on March 25 to 11:59 pm on March 27, 2018 (with a clear sky and the relative humidity of ~35%). The peak irradiation of sunlight was ~853.5 W/m$^2$. As shown in **Fig. 5A**, we placed a beaming system and a control system (with no beaming architecture, i.e., the system with a shadow board used in Fig. 3C) on the ground, ~5 meter away from the door of the house. As shown in **Fig. 5B**, three temperature curves were recorded for the ambient, the beaming system and the control system, respectively. Remarkably, the temperature in the beaming system was always lower than the ambient temperature. An obvious spike was observed on the first night due to a thin cloud on the direct top sky, reflecting the weather-dependency of radiative cooling (see ***Note S4, Figs. S4 and S5*** for another out-door experiment performed at Thuwal, Saudi Arabia with a complete different weather condition). To further reveal the cooling performance, the temperature differences in these two systems are plotted in **Fig. 5C**, showing that the beaming system reduced the temperature by 2~6 °C during the daytime and 7~9 °C at nighttime, respectively. Although the planer PDMS cooling system realized slightly better cooling performance at night (9~11 °C), its temperature is 4~6 °C higher than the surrounding in the morning since the sunlight was not always blocked by the shadow board. The V-shaped shelter structure realized all-day radiative cooling, which is highly desired in practical applications. The actual performance and further optimization will depend on the local weather condition and the optical properties of spectral selective materials, which is under investigation but beyond the scope of this work.

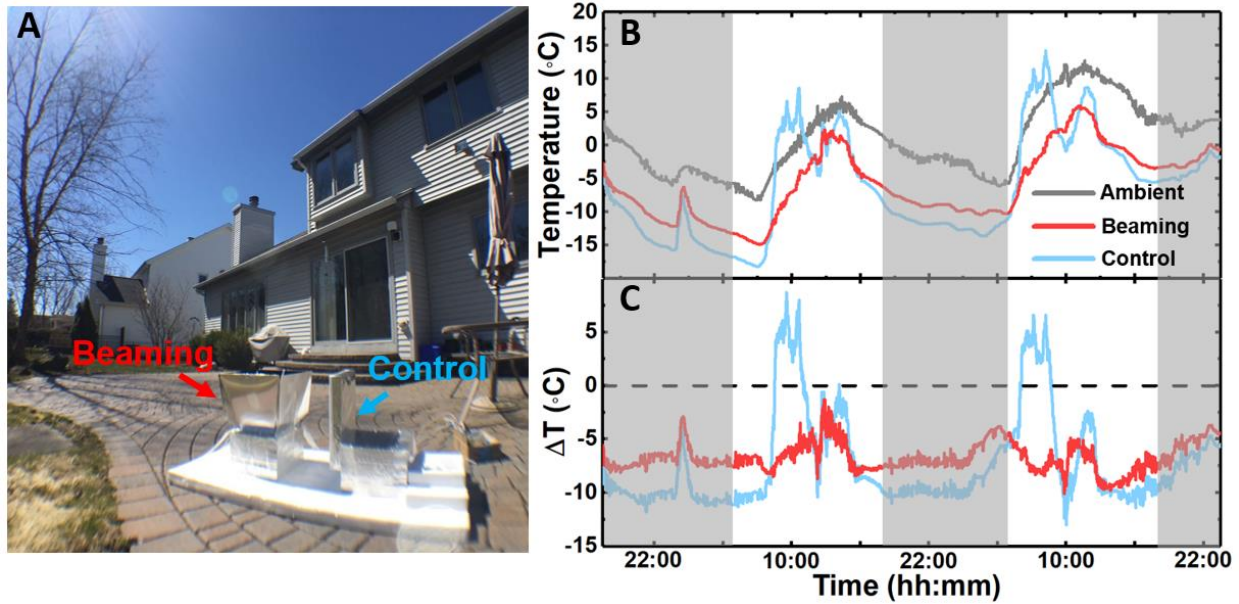

**Figure 5. All-day continuous radiative cooling.** (A) The photograph of the continuous radiative cooling experiment performed in the backyard of a house at Buffalo. (B) A continuous 50-hours cooling test: the grey line indicates the ambient temperature, the red line is the temperature in the beaming system, and the blue line is the temperature in the control system. (C) Temperature differences achieved in the beaming system (red curve) and the control system (blue curve), respectively.

In conclusion, we developed a highly efficient and low-cost passive cooling technology by exploiting the sky as the cold source. The proposed planar PDMS/Al cooling structures efficiently send invisible, heat-bearing light within the transparent window of the Earth's atmosphere (i.e., 8-13 μm) directly into the cold outer space. Using spectral selective shelter component to suppress the solar input during the daytime, this technology produced cooling effect at 2~11 °C below-ambience temperatures with the cooling power up to ~120 W/m$^2$. Importantly, such passive cooling neither consumes energy nor produces greenhouse gases. Furthermore, due to the controlled thermal emission enabled by the tapered thermal light waveguide, the beaming radiative cooling system is insensitive to the surrounding building architectures, which is therefore suitable for the implementation in urban environment. All-day continuous cooling was experimentally demonstrated on a typical sunny day at Buffalo. The large-scale production cost of the surface structure is expected to be highly competitive compared to traditional active cooling methods (e.g. electric air conditioning) because of almost zero operation cost. The proposed technology thus has disruptive potentials in transforming cooling solutions in a wide range of industrial and residential applications.

## Acknowledgements

This work was partially supported by the National Science Foundation (grand no. ECCS1507312, CBET 1445934 and ECCS 1425648).

## Author contributions

Q.G., B.O. and Z.Y conceived the idea and supervised the project. L.Z., H.S., J.L., E. S. and T. N. executed the experiments. All authors contributed to the analysis of the experimental results and modeling. L.Z., H.S., Z.Y., B.O. and Q.G. wrote the manuscript. All authors reviewed the manuscript. L.Z, H.S and J.L are co-first authors and contributed equally.

## Competing interests

Q.G. and Z.Y. have founded a company, Sunny Clean Water LLC, seeking to commercialize the results reported in this paper.